\documentclass[aps,pre,preprint,superscriptaddress]{revtex4-1}  
\usepackage{graphicx}
\usepackage{subcaption}
\usepackage{amssymb}
\usepackage{amsmath}
\usepackage{newtxtext,newtxmath,lipsum}
\usepackage[shortlabels]{enumitem}

\begin{document}
	\title{Rogue waves with two different double-periodic wave backgrounds and their modulational instabilities of a fifth-order nonlinear Schr\"odinger equation}
	\author{N. Sinthuja} 
	\affiliation{Department of Nonlinear Dynamics, Bharathidasan University, Tiruchirappalli 620024, Tamil Nadu, India}
	
	
	\author{M. Senthilvelan}
	\affiliation{Department of Nonlinear Dynamics, Bharathidasan University, Tiruchirappalli 620024, Tamil Nadu, India}
	\email{velan@cnld.bdu.ac.in}
	\begin{abstract}
		\par   In this article, we derive rogue wave (RW) solutions of a fifth-order nonlinear Schr\"odinger equation over a double-periodic wave background. Choosing the elliptic functions (combinations of $cn$, $dn$ and $sn$) as seed solutions in the first iteration of Darboux transformation and utilizing the nonlinearization of Lax pair procedure, we create the double-periodic wave background for the fifth-order nonlinear Schr\"odinger equation. By introducing the second linearly independent solution, we generate the RW solutions on the created background for three different eigenvalues. We demonstrate the differences that occur in the appearance of RWs due to the lower-order and higher-order dispersions terms. We examine the derived solution in detail for certain system and elliptic modulus parameters values and highlight some interesting features that we obtain from our studies. We also calculate the growth rate for instability of double-periodic solutions under different values of elliptic modulus parameter. 
	\end{abstract}
	
	\maketitle
	\section{Introduction}
	The nonlinear Schr\"odinger (NLS) equation is widely used to detail water waves \cite{ac}, propagation of pulses in optical fibers \cite{at} and few other aspects in physics \cite{zhakarovv,hb,ac5}. The solutions of this equation (plane wave, solitary wave, travelling wave, breathers and RW) describe various nonlinear phenomena that occur in different areas of physics. The RW phenomenon occurs in oceanography \cite{ck}, optical fiber \cite{cl}, plasmas \cite{aa}, Bose-Einstein condensates \cite{xl}, fluid mechanics \cite{sl} and Heisenberg ferromagnet \cite{bq}. In oceanography, RWs or freak waves are the waves that appear in the open sea with very high amplitudes in comparison with ambient waves \cite{ck}. In general, tsunamis and storms can be predicted hours or even days ahead of time. However, predicting oceanic RWs is impossible due to the highly erratic nature of these waves-appearing and disappearing suddenly, without a trace \cite{na}. Physical measurements have shown that these enormous waves are capable of wrecking a ship. A complete understanding of these RWs is yet to be achieved since it is both difficult and dangerous to monitor them practically. Rogue waves appear not only in the ocean but also in optics because of the modulational instability (MI) of the background waves \cite{bkibl,na1,jms,chen4,aly}. Linear models struggle to describe the high amplitude waves that occur in the open ocean. Hence, efforts have been made to investigate these RWs using nonlinear wave theory.
	
	RWs arise from a variety of backgrounds such as constant \cite{km}, non-constant wave (multi-soliton and periodic) backgrounds \cite{dj,jb,wq,xg,hq,cc,wl,bx}. A literature review showed that the RW solutions have been obtained mostly through Hirota bilinearization procedure \cite{rh,yca} and Darboux transformation (DT) method \cite{yt}. Initially considerable amount of studies have been devoted to study the RWs on constant background both at theoretical and experimental level. Very recently, constructing RWs on the periodic wave backgrounds have drawn great interest in soliton theory. This is because in nature the surface of the ocean and the waves on it shows a periodic (periodic in space) or double-periodic (periodic in both space and time) pattern.
	
	Kedziora et al., for the first time, numerically computed periodic RWs using the DT method \cite{dj}. Chen and his collaborators developed a novel method for obtaining RW solutions on a single-periodic background for systems such as NLS \cite{chen2}, modified Korteweg–de Vries (mKdV) \cite{chen1} and derivative NLS equations \cite{chen3}. Calini and Schober were reported RWs over a double-periodic background with a more accurate numerical scheme \cite{ac1} . Based on this, recently, the method of constructing RW solutions, analytically, with a double-periodic background was presented for the NLS equation \cite{chen5}. Furthermore, Pelinovsky had computed the rate of instability for both the single- and double-periodic solutions for the NLS equation \cite{dep}. Continuous RWs were experimentally observed in laser optics and water tanks \cite{rand}. The double-periodic wave solutions are also observed experimentally in hydrodynamical experiments \cite{kim}. The analytical and numerical approaches used to study the RWs over a periodic wave background have also been extended to few other higher-order NLS equations such as Hirota \cite{wq}, fourth-order NLS \cite{hqz} and fifth-order NLS equation \cite{sinthu3}. As far as higher-order NLS equations are concerned, RW over the double-periodic wave background has been constructed only for the Hirota equation. 
	
	The higher-order NLS equations have been frequently used to examine external elements such as sea depth, bottom friction, viscosity, and other parameters in ocean waves and ultra short pulse propagation in fiber optics. One such higher-order NLS is the fifth-order NLS equation. The fifth-order NLS equation has a wide range of applications and is of significant interest. From the optic and oceanographic study perspectives, the higher-order effects in solitons and RWs can be understood using the results of the fifth-order NLS equation. Research has focused on several aspects of constructing and analyzing the properties of localized solutions in the fifth-order NLS equation over the last two decades \cite{wzzhao,zwang, mal3,wrsun,hxjia,mwang}, including RWs solutions on constant wave background. Recently, we have studied the formation of RWs on the elliptic function background and investigated MI of the single-periodic waves for a fifth-order NLS equation \cite{sinthu3}. The fifth-order NLS equation, which we considered in our paper \cite{sinthu3}, is derived from the deformation of the inhomogeneous Heisenberg ferromagnetic system with the help of a prolongation structure. Very recently, Wang and Zhaqilao have studied RWs solutions of another fifth-order NLS equation on two different single-periodic wave backgrounds and reported how the amplitude and background waves of the RW solutions change with respect to the higher-order dispersion terms \cite{zwang}. In this present work, we aim to construct RW solutions over the double-periodic wave background and study their MI for the fifth-order NLS equation \cite{Feng,song}.
	
	The following explains the structure of our paper: In Section 2, we present a fifth-order NLS equation and its Lax pair and DT applicable for this equation. In Section 3, we demonstrate the underlying procedure, namley nonlinearization of Lax pair for creating background waves. In Section 4, we introduce a second solution of Lax pair equation to generate the RWs over double-periodic background waves. In Section 5, we compute the instability rate for the seed solution, which are double-periodic wave solutions of fifth-order NLS equation. The summary of our work is described in Section 6.
	\section{Darboux transformation and double-periodic solutions of Eq. (\ref{e1})}
	In the present research work, we consider a fifth-order NLS equation in the following form,
	\begin{align}
		iq_t+\frac{1}{2}q_{xx}+|q|^2q-i\delta(q_{xxxxx}+10|q|^2q_{xxx}&+20\bar{q}q_xq_{xx}+30|q|^4q_x\nonumber\\&+10(q|q_x|^2)_x)=0,
		\label{e1} 
	\end{align} 
	in which the function $\bar{q}=\bar{q}(x,t)$ is a complex conjugation of $q=q(x,t)$ while the function denotes wave envelope. The real parameter $\delta$ can be used independently to control the values of the fifth-order dispersion and to learn the changes that occur in the dynamics of the system (\ref{e1}). 
	The variables $x$ and $t$ are placed in the subscripts of $q$ denote the partial derivative with respect to spatial and temporal coordinates, respectively.
	
	The existence of Lax pair (system of linear equation) for the Eq. (\ref{e1}) can be read from the following
	\begin{subequations}
		\label{s2}
		\begin{align}
			&	\varphi_{x}=L(\lambda,q) \varphi,\;\; \varphi_{t}  = M(\lambda,q) \varphi, \;\; \label{e2}\\ &L(\lambda,q)=\begin{pmatrix} \lambda & q\\ -\bar{q} & -\lambda \end{pmatrix},\;\; M(\lambda,q)=\begin{pmatrix} A_1  & A_2\\ A_3 & -A_1 \end{pmatrix},\;\;
			\label{e122}	  
		\end{align}
	\end{subequations}
	with
	\begin{align}
		A_1=\;&i\left(\frac{|q|^2}{2}+\lambda^2\right)+\delta(\bar{q}q_{xxx}-q\bar{q}_{xxx}+q_x\bar{q}_{xx}-\bar{q_x}q_{xx}-6|q|^2(q\bar{q}_x-\bar{q}q_x))\nonumber\\&+2\lambda\delta(\bar{q}q_{xx}+q\bar{q}_{xx}-|q_x|^2+3|q|^4)-4\lambda^2\delta(q\bar{q}_x-\bar{q}q_x)+8\lambda^3\delta|q|^2+16\lambda^5\delta,\nonumber\\
		A_2=&\;i\left(\frac{q_x}{2}+\lambda q\right)+\delta(q_{xxxx}+8|q|^2q_{xxx}+2q^2\bar{q}_{xx}+4|q_x|^2q+6\bar{q}q^2_x+6|q|^4q)\nonumber\\&+2\lambda\delta(q_{xxx}+6|q|^2q_x)+4\lambda^2\delta(q_{xx}+2|q|^2q)+8\lambda^3\delta q_x+16\lambda^4\delta q,\nonumber\\
		A_3=&\;i\left(\frac{\bar{q}_x}{2}-\lambda\bar{q}\right)-\delta(\bar{q}_{xxxx}+8|q|^2\bar{q}_{xxx}+2\bar{q}^2q_{xx}+4|q_x|^2\bar{q}+6q\bar{q}^2_x+6|q|^4\bar{q})\nonumber\\&+2\lambda\delta(\bar{q}_{xxx}+6|q|^2\bar{q}_x)-4\lambda^2\delta(\bar{q}_{xx}+2|q|^2\bar{q})+8\lambda^3\delta\bar{q}_x-16\lambda^4\delta\bar{q}.\nonumber
	\end{align}
	
	In the above, the function $\varphi=(\psi_1,\varphi_1)^{T}$ is the solution of the Lax pair Eq. (\ref{s2}) wherein the superscript $T$ stands for the vector transpose. The spectral parameter is denoted by the Greek alphabet $\lambda$. One can unambiguously prove that the zero curvature condition, $L_t-M_x+LM-ML=0$, commensurate to Eq. (\ref{e1}).
	
	The one-fold DT of Eq. (\ref{e1}) is given by \cite{mwang}
	\begin{equation}
		\hat{q}(x,t)=q(x,t)+\frac{2(\lambda_1+\bar{\lambda}_1)\varphi_1 \bar{\varphi}_2}{|\varphi_1|^2+|\varphi_2|^2},
		\label{e3}
	\end{equation}
	where $\hat{q}(x,t)$ and $q(x,t)$ are the first iterated and seed solution of Eq. (\ref{s2}). One can solve the Lax pair Eq. (\ref{s2}) with various forms of seed solutions. For example, the seed solution in the form of plane and periodic waves yield RWs over the respective background. On the other hand the choice $q(x,t)=0$ helps one to identify soliton solutions for the Eq. (\ref{e1}). Even among the periodic waves one may consider various forms to feed as seed solution. For example, one may consider the periodicity in only one direction (spatial) or in two directions (spatial and temporal) and the respective expression can be fed as a seed solution in Eq. (\ref{s2}). From a mathematical standpoint, one can observe that it becomes less tedious to solve the linear Eq. (\ref{s2}) using seed solutions that are periodic in only one direction when compared to using seed solutions that are periodic in both space and time. In the following, we solve the Lax pair Eq. (\ref{s2}) with double-periodic wave solutions (waves that are periodic in space and time) as seed solution.
	\begin{figure*}[!ht]
		\begin{center}
			\begin{subfigure}{0.4\textwidth}
				\includegraphics[width=\linewidth]{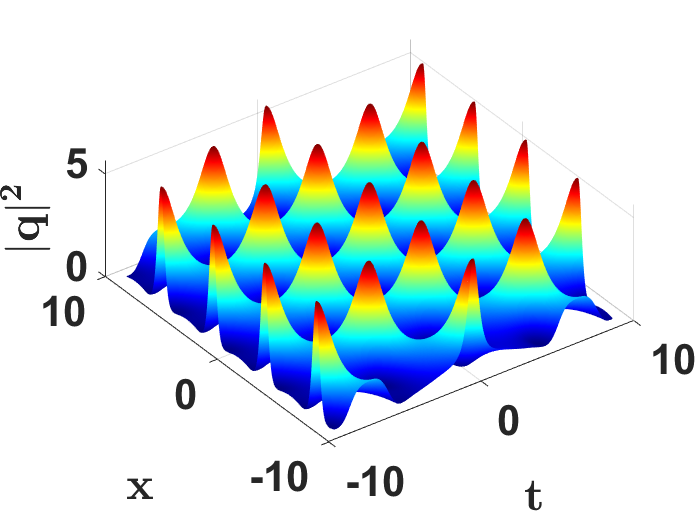}
				\caption{}
			\end{subfigure}
			\begin{subfigure}{0.4\textwidth}
				\includegraphics[width=\linewidth]{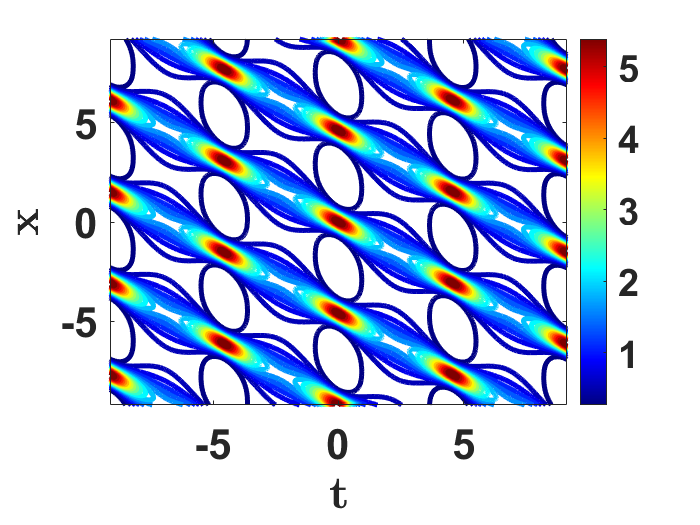}
				\caption{}
			\end{subfigure}\\
			\begin{subfigure}{0.4\textwidth}
				\includegraphics[width=\linewidth]{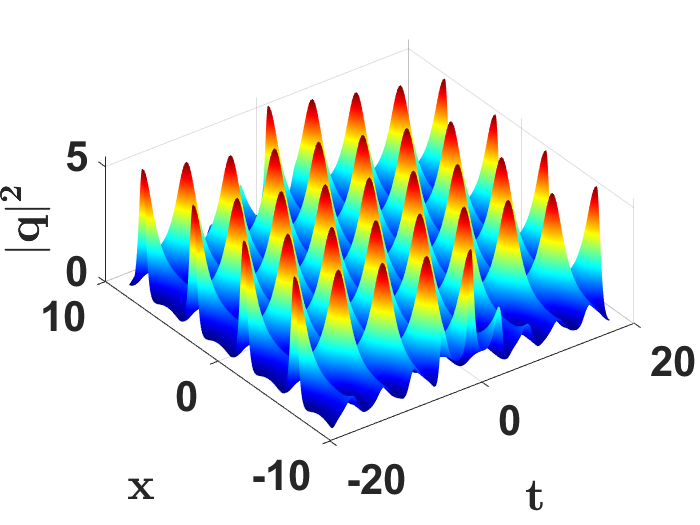}
				\caption{}
			\end{subfigure}
			\begin{subfigure}{0.4\textwidth}
				\includegraphics[width=\linewidth]{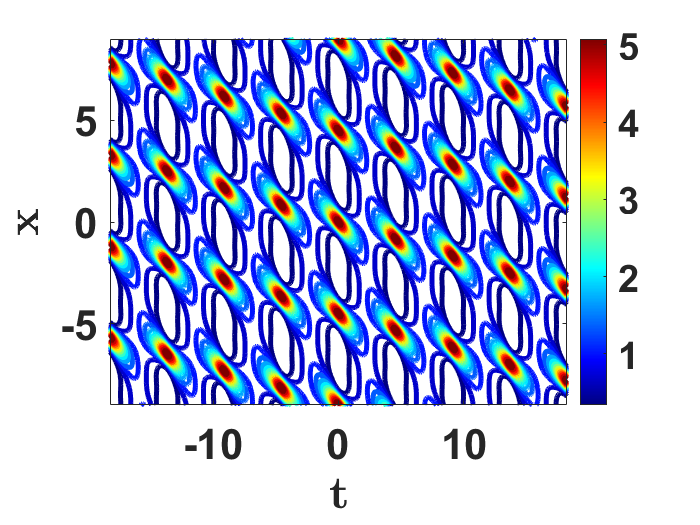}
				\caption{}
			\end{subfigure}\\
			\begin{subfigure}{0.4\textwidth}
				\includegraphics[width=\linewidth]{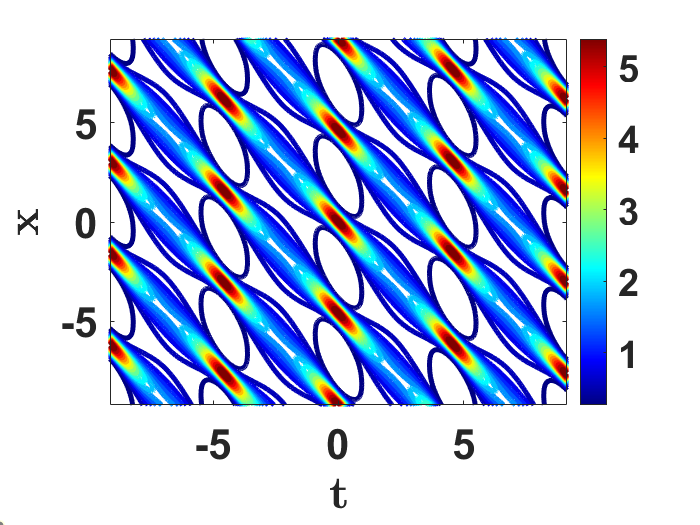}
				\caption{}
			\end{subfigure}
			\begin{subfigure}{0.4\textwidth}
				\includegraphics[width=\linewidth]{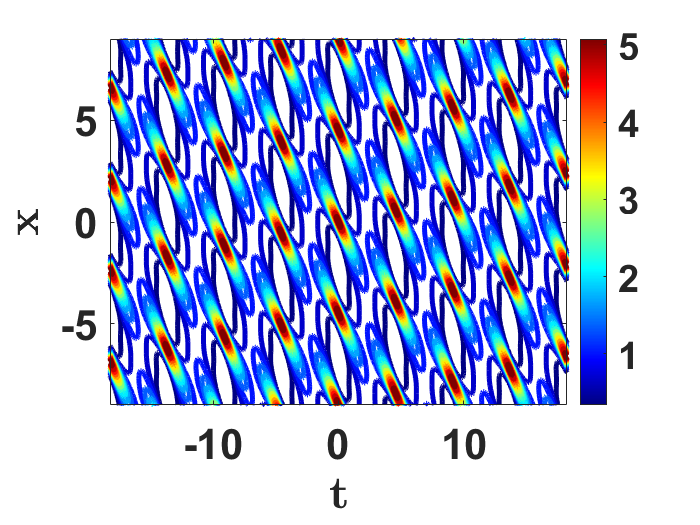}
				\caption{}
			\end{subfigure}
		\end{center}
		\vspace{-0.3cm}
		\caption{ The wave pattern profiles of double-periodic waves for $k=0.9$: (a)-(b) The phase-repeated wave patterns and its corresponding contour plot of Eq. (\ref{e4}) with the system parameter $\delta=0.05$, (c)-(d) The phase-alternated wave patterns and its corresponding contour plot of Eq. (\ref{e5}) for $\delta=0.05$, (e)-(f) are contour plots of (\ref{e4}) and (\ref{e5}) drawn for $\delta=0.1$.}
		\label{fig1}
	\end{figure*}
	To achieve our task we proceed as follows. Instead of solving the Lax pair Eq. (\ref{s2}) with doubly periodic solution as seed solution we consider two different double-periodic traveling wave solutions of Eq. (\ref{e1}), having Jacobian elliptic functions \cite{crabb}, in the form
	\begin{align}
		q(x,t)=&k\frac{\text{cn}(t,k)~\text{cn}(\sqrt{\epsilon}(x+\omega t),\tau)+i\sqrt{\epsilon}~\text{sn}(t,k)~\text{dn}(\sqrt{\epsilon}(x+\omega t),\tau)}{\sqrt{\epsilon}~\text{dn}(\sqrt{\epsilon}(x+\omega t),\tau)-\text{dn}( t,k)~\text{cn}(\sqrt{\epsilon}(x+\omega t),\tau)}e^{i t},\quad \tau=\frac{\sqrt{1-k}}{\sqrt{\epsilon}},
		\label{e4}\\
		q(x,t)=&\frac{\text{dn}(t,k)~\text{cn}(\sqrt{2}(x+\omega t),\tau)+i\sqrt{\epsilon k}~\text{sn}(t,k)}{\sqrt{\epsilon}-\sqrt{k}~\text{cn}(t,k)~\text{cn}(\sqrt{2}(x+\omega t),\tau)}e^{i k t}, \quad  \tau=\frac{\sqrt{1-k}}{\sqrt{2}},
		\label{e5}
	\end{align}
	and determine the associated eigenvalues so that they become solutions of the Lax pair Eq. (\ref{s2}). In Eqs. (\ref{e4}) and (\ref{e5}) the elliptic modulus parameter $k$ varies from $0$ to $1$, $\epsilon=1+k$ and the constant $\omega$ is related to the system parameter $\delta$ through the relation $\omega=\left(16-\displaystyle \frac{2}{k^2}\right)\delta$. 
	
	Figure \ref{fig1} shows the unique characteristics of the double-periodic waves (\ref{e4}) and (\ref{e5}) for $k=0.9$ with two different values of system parameter ($\delta$). Figures \ref{fig1}(a)-\ref{fig1}(b) represent the three dimensional wave profile and its contour plot of Eq. (\ref{e4}) when $\delta=0.05$ respectively. Similarly for Eq. (\ref{e5}), the demonstrations are given in Figures \ref{fig1}(c) and \ref{fig1}(d). The contour plots of Eqs. (\ref{e4}) and (\ref{e5}) with $\delta=0.1$ are shown in Figures \ref{fig1}(e) and \ref{fig1}(f). Clearly, the double-periodic wave solutions (\ref{e4}) and (\ref{e5}) generate two different wave patterns, namely phase-repeated and phase-alternated wave patterns. Both the wave patterns are clearly shown in Figures \ref{fig1}(b) and \ref{fig1}(d). As the system parameter $(\delta)$ is increased from $0.05$ to $1$, the localization of waves or peaks increase in the $(x-t)$ plane when compared to lower values of $\delta(\leq0.05$). We also notice that while the system parameter ($\delta$) is varied, the orientation and the number of occurrences of a repeating wave pattern of both the waves are slightly change as illustrated in Figures \ref{fig1}(e) and \ref{fig1}(f).
	
	\section{Construction of two different double-periodic wave backgrounds  }
	
	In this section, we create double-periodic wave backgrounds for the Eq. (\ref{e1}) through DT formula (\ref{e3}). In the formula for one-fold DT (\ref{e3}), the seed solution $q(x,t)$ is considered to be a double-periodic wave and our task is to find its eigenvalues and eigenfunctions. To locate the eigenvalues of solutions (\ref{e4}) and (\ref{e5}), we use the nonlinearization of Lax pairs approach \cite{sinthu3,zhou}. In this approach, one can construct the solution from the knowledge of the squared eigenfunctions of Eq. (\ref{e2}) through the Bargmann constraint 
	\begin{align}
		q(x,t)=\psi^2_1+\psi^2_2+\bar{\varphi}^2_1+\bar{\varphi}^2_2,
		\label{e6}
	\end{align}
	where $\psi_1$, $\psi_2$, $\bar{\varphi}_1$ and $\bar{\varphi}_2$ are the solutions of the Lax pair (\ref{s2}) at $\lambda=\lambda_1$ and $\lambda=\lambda_2$.
	
	The method of constructing $q(x,t)$ has already been discussed at length in the earlier works \cite{chen5}. Hence, we have skipped over those details and provide only the final expressions. Implementing the procedure, and after a very length calculation, we obtain the squared eigenfunctions $\psi^2_1$ and $\varphi^2_1$ in the form
	\begin{align}
		\label{e7}
		\psi_1^2= & \frac{\lambda_1\left(q''+2|q|^2 q+4(b+\lambda_1^2)q+2\lambda_1 q'\right)}{4(\lambda_1+\bar{\lambda}_1)(\lambda_1+\bar{\lambda}_2)(\lambda_1-\lambda_2)},\nonumber\\
		\varphi_1^2= & \frac{\lambda_1\left(\bar{q}''+2|q|^2\bar{q}+4(b+\lambda_1^2)\bar{q}-2\lambda_1 \bar{q}'\right)}{4(\lambda_1+\bar{\lambda}_1)(\lambda_1+\bar{\lambda}_2)(\lambda_1-\lambda_2)},\nonumber\\
		\psi_1\varphi_1= & -\frac{\lambda_1\left(q'\bar{q}-q\bar{q}'+2\lambda_1 (2b+2\lambda_1^2+|q|^2)\right)}{4(\lambda_1+\bar{\lambda}_1)(\lambda_1+\bar{\lambda}_2)(\lambda_1-\lambda_2)}.
	\end{align}
	\begin{figure*}[!ht]
		\begin{center}
			\begin{subfigure}{0.4\textwidth}
				\includegraphics[width=\linewidth]{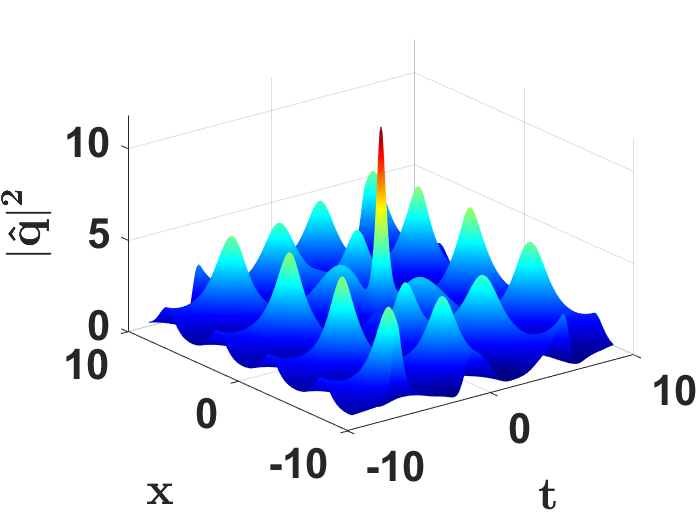}
				\caption{}
			\end{subfigure}
			\begin{subfigure}{0.4\textwidth}
				\includegraphics[width=\linewidth]{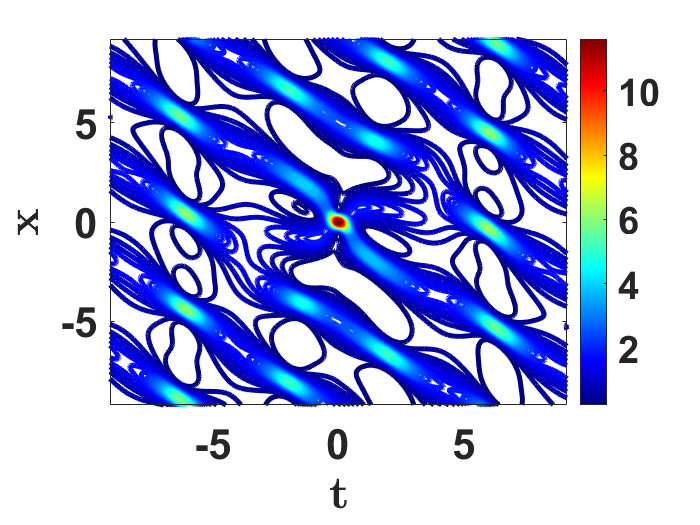}
				\caption{}
			\end{subfigure}\\
			\begin{subfigure}{0.4\textwidth}
				\includegraphics[width=\linewidth]{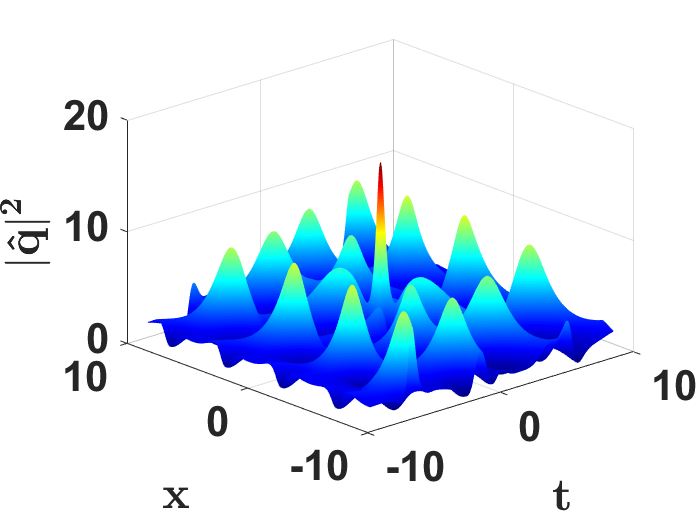}
				\caption{}
			\end{subfigure}
			\begin{subfigure}{0.4\textwidth}
				\includegraphics[width=\linewidth]{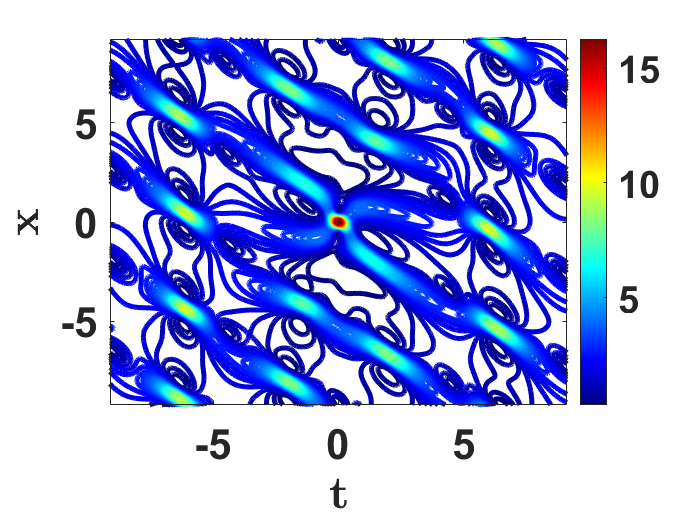}
				\caption{}
			\end{subfigure}\\
			\begin{subfigure}{0.4\textwidth}
				\includegraphics[width=\linewidth]{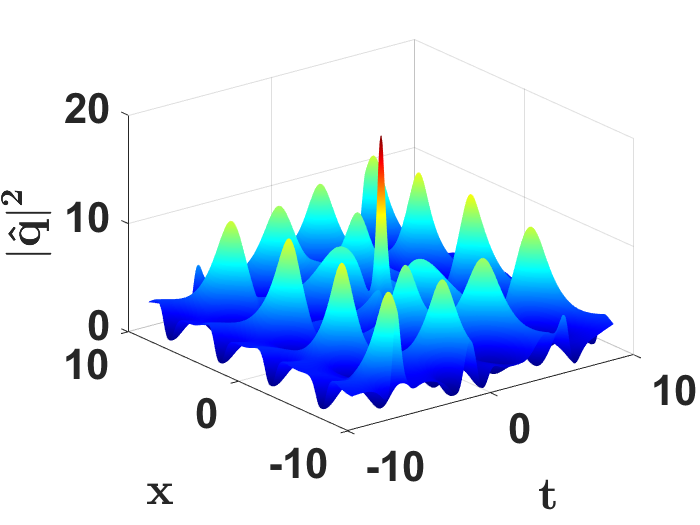}
				\caption{}
			\end{subfigure}
			\begin{subfigure}{0.4\textwidth}
				\includegraphics[width=\linewidth]{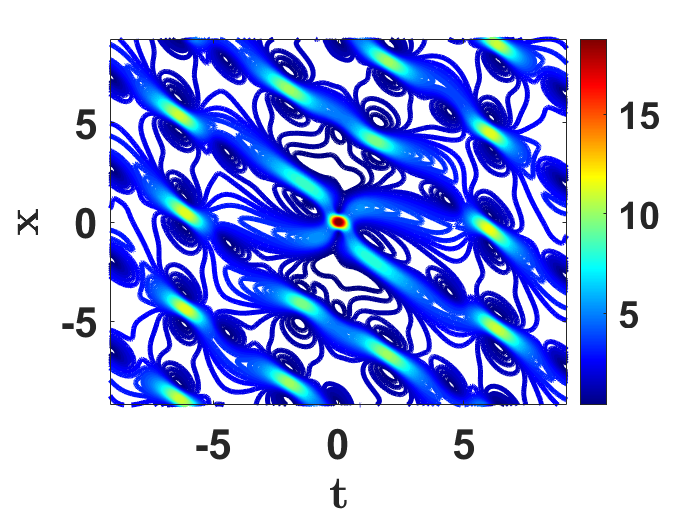}
				\caption{}
			\end{subfigure}
		\end{center}
		\vspace{-0.15cm}
		\caption{The qualitative profiles of rogue waves over a double-periodic solutions (\ref{e4}) and corresponding contour plots  for Eq. (\ref{e1}) with $k=0.9$ and $\delta=0.06$. (a)-(b) $\lambda_1=\sqrt{\nu_1}$, (c)-(d) $\lambda_2=\sqrt{\nu_2}$ and (e)-(f) $\lambda_3=\sqrt{\nu_3}$. }
		\label{fig2}
	\end{figure*}
	

	
	
	For the double-periodic wave solution Eq. (\ref{e4}), the eigenvalues are found to be real and they read
	\begin{align}
		\lambda_1=\pm\sqrt{\nu_1},\quad
		\lambda_2=\pm\sqrt{\nu_2},\quad
		\lambda_3=\pm\sqrt{\nu_3},
		\label{ee1}
	\end{align}
	where $\nu_2=\nu_3-\nu_1$, $\nu_3=1$ and the parameter $\nu_1$ varies from zero to one. 
	\begin{figure*}[!ht]
		\begin{center}
			\begin{subfigure}{0.4\textwidth}
				\includegraphics[width=\linewidth]{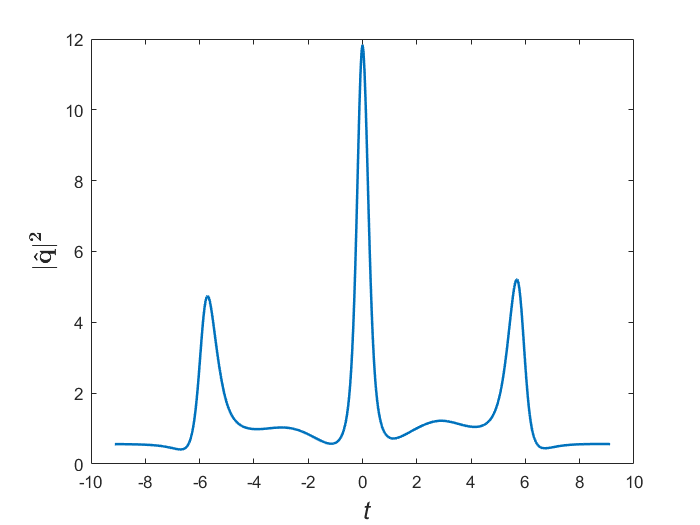}
				\caption{}
			\end{subfigure}
			\begin{subfigure}{0.4\textwidth}
				\includegraphics[width=\linewidth]{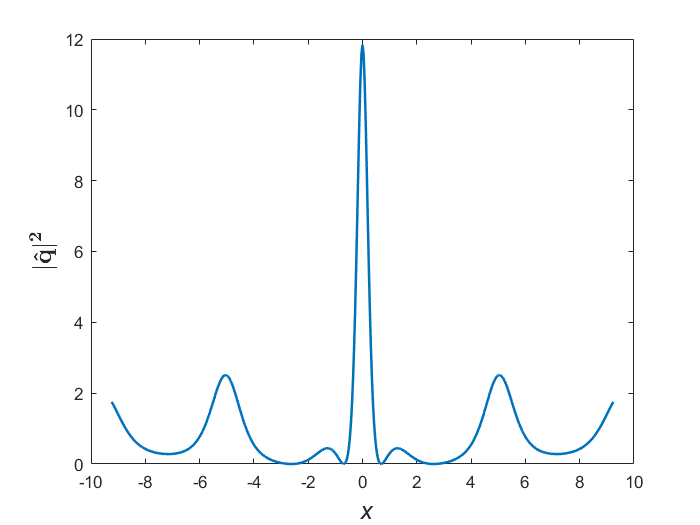}
				\caption{}
			\end{subfigure}\\
			\begin{subfigure}{0.4\textwidth}
				\includegraphics[width=\linewidth]{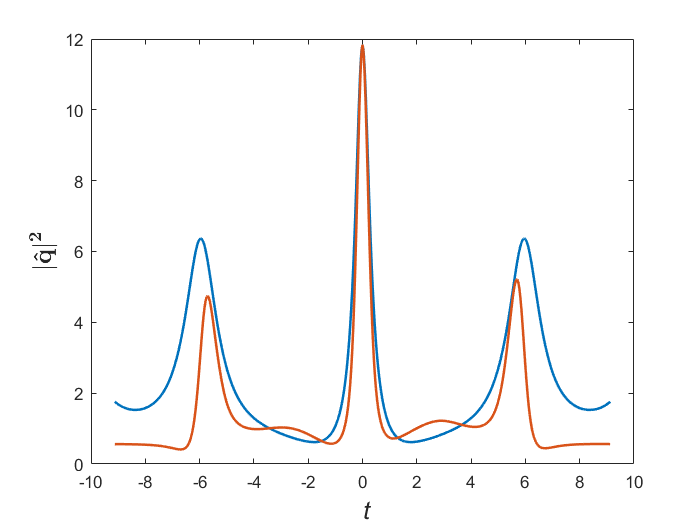}
				\caption{}
			\end{subfigure}
			\begin{subfigure}{0.4\textwidth}
				\includegraphics[width=\linewidth]{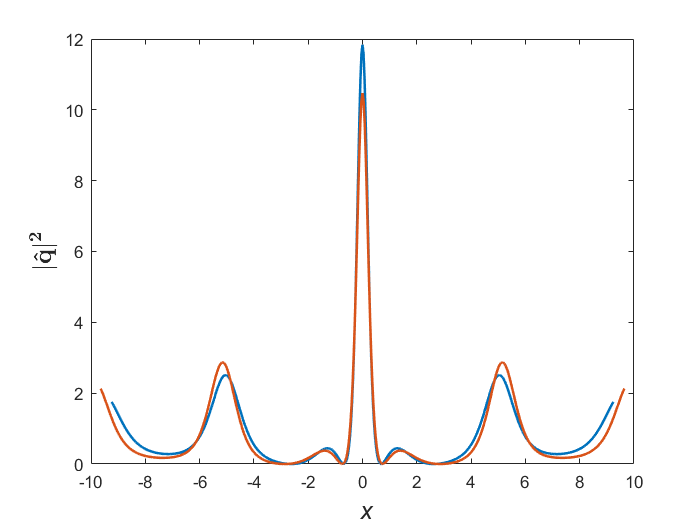}
				\caption{}
			\end{subfigure}
		\end{center}
		\vspace{-0.15cm}
		\caption{Two dimensional plots for rogue waves over a double-periodic solution (\ref{e4}) of Eq. (\ref{e1}) with $\lambda_1=\sqrt{\nu_1}$. (a) $x=0$,$k=0.9$ and $\delta=0.06$, (b) $t=0$, $k=0.9$ and $\delta=0.06$, (c) $x=0$, $k=0.9$ and two different values of $\delta$ ($0$ [blue color], $0.06$ [orange color]) and (d) $t=0$, $\delta=0.06$ and two different values of $k$ ($0.8$ [orange color],$0.9$ [blue color]).}
		\label{fig3}
	\end{figure*}

	In the case of double-periodic wave solution Eq. (\ref{e5}), we found only one eigenvalue as real and the other two turned out to be complex conjugates, that is
	\begin{align} 
		\lambda_1=\pm\sqrt{\nu_1},\quad
		\lambda_2=\pm\sqrt{\alpha+i\beta},\quad \lambda_3=\pm\sqrt{\alpha-i\beta},
		\label{ee2}
	\end{align}
	where $\alpha=\frac{\nu_1}{2}$ and $\beta=\sqrt{(\frac{1}{2}-\alpha)(\frac{1}{2}+\alpha)}$.  
	
	Once the eigenvalues and squared eigenfunctions are identified, by substituting these expressions and the double-periodic wave solutions in the onefold-DT (\ref{e3}) we can generate the waves that are in the form of periodic in both space and time and which in turn constitute the background waves for the fifth-order NLS Eq. (\ref{e1}). Our next task is to engender RWs over these double-periodic wave backgrounds. For this purpose, we proceed to establish another solution of Eq. (\ref{s2}), which is a linearly independent one.
	\begin{figure*}[!ht]
		\begin{center}
			\begin{subfigure}{0.4\textwidth}
				\includegraphics[width=\linewidth]{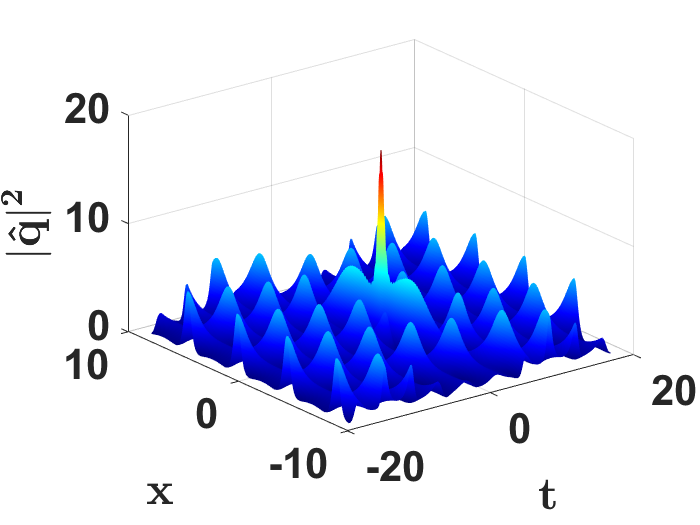}
				\caption{}
			\end{subfigure}
			\begin{subfigure}{0.4\textwidth}
				\includegraphics[width=\linewidth]{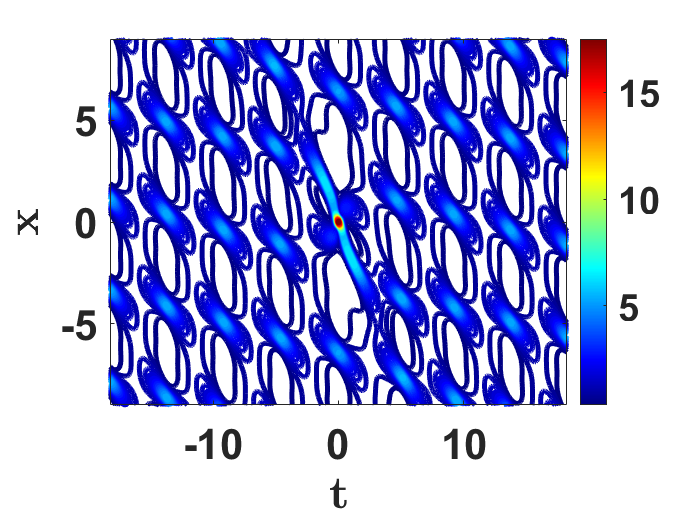}
				\caption{}
			\end{subfigure}\\
			\begin{subfigure}{0.4\textwidth}
				\includegraphics[width=\linewidth]{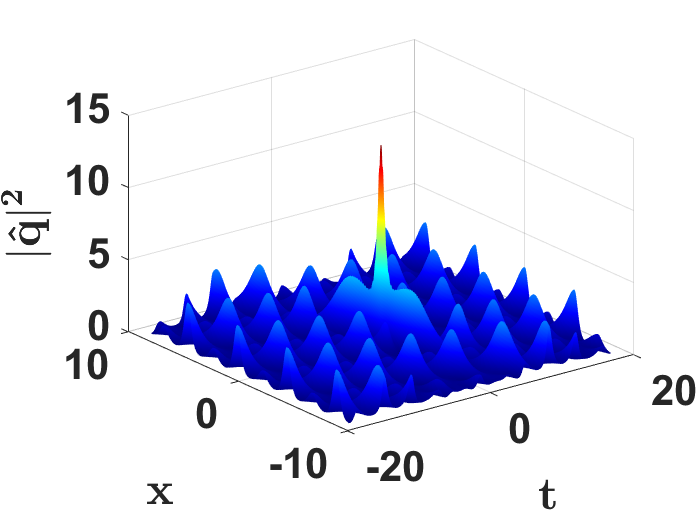}
				\caption{}
			\end{subfigure}
			\begin{subfigure}{0.4\textwidth}
				\includegraphics[width=\linewidth]{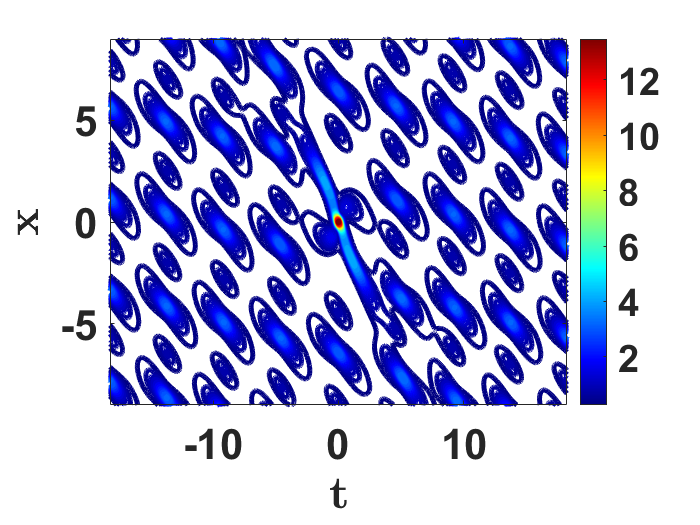}
				\caption{}
			\end{subfigure}
		\end{center}
		\vspace{-0.15cm}
		\caption{The qualitative profiles of rogue waves over a double-periodic solutions Eq. (\ref{e5}) and corresponding contour plots  for Eq. (\ref{e1}) with $k=0.9$ and $\delta=0.05$. (a)-(b) $\lambda_1=\sqrt{\nu_1}$, (c)-(d) $\lambda_2=\sqrt{\alpha+i\beta}$.}
		\label{fig4}
	\end{figure*}
	\section{Construction of RW solutions on a double-periodic wave background}
	
	We choose the second solution $\varphi=(\hat{\psi}_1,\hat{\varphi}_1)^{T}$ of the linear Eq. (\ref{s2}), as is given by \cite{chen5}
	\begin{align}
		\label{e8}
		\hat{\psi}_1 = \psi_1 \gamma_1-\frac{2\bar{\varphi}_1}{|\psi_1|^2+|\varphi_1|^2}, \quad \hat{\varphi}_1 =  \varphi_1 \gamma_1+\frac{2\bar{\psi}_1}{|\psi_1|^2+|\varphi_1|^2},
	\end{align}
	where $\gamma_1$ is an unknown function that must be determined. 
	
	The explicit expression for the unknown function $\gamma_1$ can be determined from Eq. (\ref{s2}). 
	Substituting Eq. (\ref{e8}) into the first equation in Eq. (\ref{s2}), we obtain
	\begin{align}
		\label{e10}	
		\gamma_{1x} = D_1:= & -\frac{4(\lambda_1+\bar{\lambda}_1)\bar{\psi}_1\bar{\varphi}_1}{\left(|\hat{\psi}_1|^2+|\hat{\varphi}_1|^2\right)^2}.
	\end{align}
	Inserting Eq. (\ref{e8}) into the second equation in Eq. (\ref{s2}), we found
	\begin{align}
		\gamma_{1t} = D_2: = &\frac{2 (\lambda_1+\bar{\lambda}_1)S_1{\bar{\psi}_1}^2}{(|\psi_1|^2+|\varphi_1|^2)^2}
		+\frac{2 (\lambda_1+\bar{\lambda}_1)S_2\bar{\varphi}_1^2}{(|\psi_1|^2+|\varphi_1|^2)^2}-\frac{4 (\lambda_1+\bar{\lambda}_1)S_3\bar{\psi}_1\bar{\varphi}_1}{(|\psi_1|^2+|\varphi_1|^2)^2},
		\label{s3}
	\end{align}
	where
	\begin{align}
		S_1=~&8\delta(\lambda_1-\bar{\lambda}_1)|q|^2q+q(i+12\delta\bar{q}q_x+16\delta(\lambda_1-\bar{\lambda}_1)(\lambda_1^2+{\bar{\lambda}_1}^2))+2\delta(q_{xxx}\nonumber\\&+2(\lambda_1-\bar{\lambda}_1)q_{xx}+4(\lambda^2_1-|\lambda_1|^2+\bar{\lambda}_1^2)q_x),\nonumber\\
		S_2=~&8\delta(\lambda_1-\bar{\lambda}_1)|q|^2\bar{q}+\bar{q}(i-12\delta q\bar{q}_x+16\delta(\lambda_1-\bar{\lambda}_1)(\lambda_1^2+{\bar{\lambda}_1}^2))-2\delta(\bar{q}_{xxx}\nonumber\\&-2(\lambda_1-\bar{\lambda}_1)\bar{q}_{xx}+4(\lambda^2_1-|\lambda_1|^2+\bar{\lambda}_1^2)\bar{q}_x),\nonumber
	\end{align}
	\begin{align}
		S_3=~& i(\lambda_1-\bar{\lambda}_1)+2\delta(3|q|^4+\bar{q}q_{xx}-|q_x|^2+2(\lambda_1-\bar{\lambda}_1)\bar{q}q_x+8(\lambda_1^4-|\lambda_1|^2\times\nonumber\\&(\lambda_1^2+\bar{\lambda}_1^2)+|\lambda_1|^4+\bar{\lambda}_1^4)+q(\bar{q}_{xx}-2(\lambda_1-\bar{\lambda}_1)\bar{q}_x+4(\lambda_1^2-|\lambda_1|^2+\bar{\lambda}_1^2)\bar{q})).\nonumber
	\end{align}
	The system of first-order partial differential Eqs. (\ref{e10}) and (\ref{s3}) are compatible with each other through this relation $D_{1t}=D_{2x}$. Upon integrating Eqs. (\ref{e10}) and (\ref{s3}), we get 
	\begin{align}
		\label{e12}
		\gamma_1(x,t)= & \int_{x_0}^{x} D_1(x',t_0)dx'+ \int_{t_0}^t D_2(x_0,t')dt',
	\end{align}
	where $x_0$ and $t_0$ are fixed arbitrarily. Since the functions $D_1(x,t)$ and $D_2(x,t)$ have complicated expressions we move on to solve the above Eq. (\ref{e12}), numerically.
	
	Replacing the second linearly independent solutions $\hat{\psi}_1$ and $\hat{\varphi}_1$, instead of the first solutions $\psi_1$ and $\varphi_1$, in the DT formula (\ref{e3}), we found
	\begin{align}
		\label{e13}
		\hat{q}(x,t)= & q(x,t)+\frac{2(\lambda_1+\bar{\lambda}_1)\hat{\psi}_1\bar{\hat{\varphi}}_1}{|\hat{\psi}_1|^2+|\hat{\varphi}_1|^2},  
	\end{align}
	in which $\hat{\psi}_1$ and $\hat{\varphi}_1$ are given in Eq. (\ref{e8}). Equation (\ref{e13}) yields a novel solution for the fifth-order NLS Eq. (\ref{e1}). 
	
	We have two families of double-periodic solutions Eqs. (\ref{e4}) and (\ref{e5}), each has a set of three eigenvalues. By substituting the double periodic solution in Eq. (\ref{e4}) with the first eigenvalue $\lambda_1=\pm\sqrt{\nu_1}$, squared eigenfunctions and also the second solution Eq. (\ref{e8}) in Eq. (\ref{e13}), we can generate the RW over a double-periodic wave background. By replacing the eigenvalues $\lambda_2$ and $\lambda_3$ separately in place of $\lambda_1$, we came across RWs with different amplitudes on the same doubly periodic wave background. A similar action with Eq. (\ref{e5}) also yield RWs of different nature.  
	
	Figure \ref{fig2} represents the three-dimensional plots of $|\hat{q}|^2$ and their corresponding contour plots of RW over the double-periodic solution Eq. (\ref{e4}) of the fifth-order NLS Eq. (\ref{e1}) for $k=0.9$, $\delta=0.06$ and three different sets of real eigenvalues (\ref{ee1}). The RWs obtained with the background of double-periodic wave for $\lambda_1=\sqrt{\nu_1}$ ($\nu_1$ is the free parameter, which varies from $0$ to $0.5$) is illustrated in Figure \ref{fig2}(a) and its contour plot is given in Figure \ref{fig2}(b). Figures \ref{fig2}(c) and \ref{fig2}(d) which are same as Figs. \ref{fig2}(a) and \ref{fig2}(b) but drawn for $\lambda_2=\sqrt{\nu_2}$, where $\nu_2=\nu_3-\nu_1$ and $\nu_3=1$. The outcome for the eigenvalue $\lambda_3=\sqrt{\nu_3}$ is produced in Figures \ref{fig2}(e)-\ref{fig2}(f). In all the cases, we notice that the amplitude of RWs attain their highest value at the origin $(x,t)=(0,0)$. However, they differ from each other in amplitude. This is because there is a magnitude of variation in the eigenvalues. The magnitude of eigenvalues depends on the elliptic modulus parameter in the form $k=2\sqrt{\nu_1 \nu_2}$. The amplitude $|\hat{q}|^2$ varies in each case $( 11.8376<16.5893<19.1704)$ in the order of eigenvalues $\lambda_1<\lambda_2<\lambda_3$. We could also visualize changes in their orientation (see their respective contour plots).
	\begin{figure*}[!ht]
		\begin{center}
			\begin{subfigure}{0.4\textwidth}
				\includegraphics[width=\linewidth]{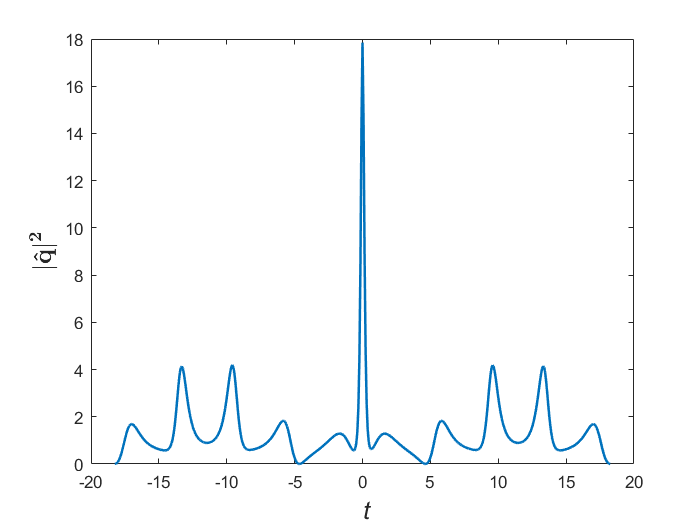}
				\caption{}
			\end{subfigure}
			\begin{subfigure}{0.4\textwidth}
				\includegraphics[width=\linewidth]{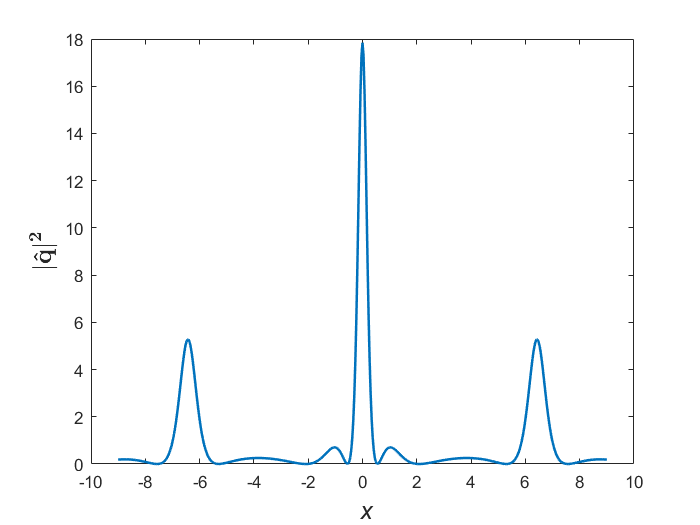}
				\caption{}
			\end{subfigure}\\
			\begin{subfigure}{0.4\textwidth}
				\includegraphics[width=\linewidth]{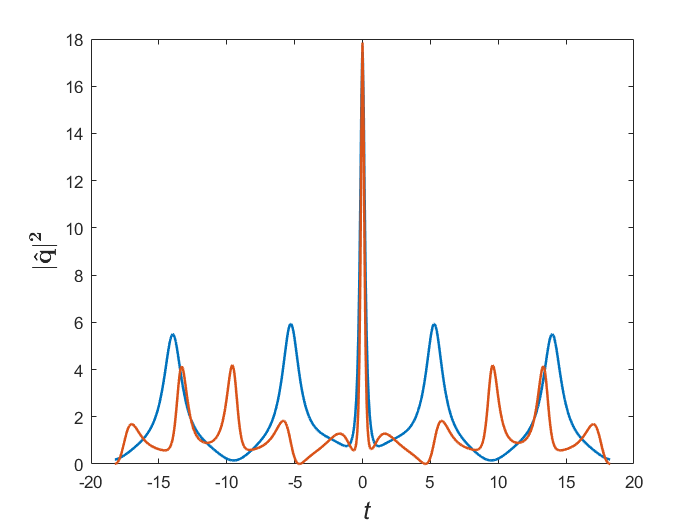}
				\caption{}
			\end{subfigure}
			\begin{subfigure}{0.4\textwidth}
				\includegraphics[width=\linewidth]{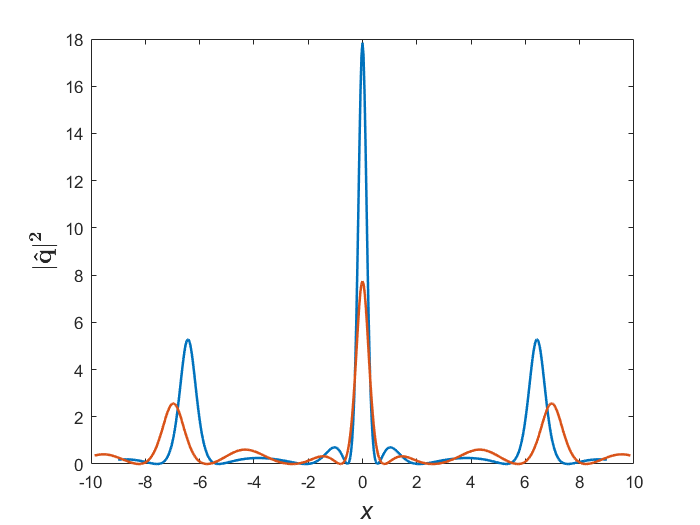}
				\caption{}
			\end{subfigure}
		\end{center}
		\vspace{-0.15cm}
		\caption{Two dimensional plots for rogue waves over a double-periodic solution (\ref{e5}) of Eq. (\ref{e1}) with $\lambda_1=\sqrt{\nu_1}$. (a) $x=0$, $k=0.9$ and $\delta=0.05$, (b) $t=0$, $k=0.9$ and $\delta=0.05$, (c) $x=0$, $k=0.9$ and two different values of $\delta$ ( $0$ [blue color], $0.05$ [orange color]) and (d) $t=0$, $\delta=0.05$ and two different values of $k$ ($0.3$ [orange color],$0.9$ [blue color]).}
		\label{fig5}
	\end{figure*}
	
	Two dimensional plots of $|\hat{q}|^2$ on the same background (\ref{e4}) with $\lambda_1=\sqrt{\nu_1}$ are given in Figure \ref{fig3}. For $k=0.9$ and $\delta=0.06$, the cross sectional RW pattern obtained in $t$-plane at $x=0$ is displayed in Figure \ref{fig3}(a) and a similar cross sectional wave pattern was observed in the $x$ plane at $t=0$ (Figure \ref{fig3}(b)). In both the cross sectional views, as one can see, the amplitude of the RWs attain their maximum ($\approx 12$) with different amplitudes of the background waves. In Figure \ref{fig3}(c), the RWs with double-periodic wave solution of Eq. (\ref{e1}) with $x=0$, $k=0.9$ for different values of system parameter $(\delta)$ is shown. The blue color peaks represent the two dimensional RW pattern of Eq. (\ref{e1}) with $\delta=0$, that is, it represents the RW solution in the respective background of the standard NLS equation, whereas the orange color peaks denote the RWs for $\delta=0.06$. When we vary the system parameter from $0$ to $0.06$, we observe a couple of differences that occur (Figure \ref{fig3}(c)). Firstly, the background waves amplitude decreases and there are changes in the orientation (not very much high) in the $t$-plane. Secondly, RW becomes sharp whose width narrows to lie inside the width of NLS RW. When $\delta=0.06$ and $t=0$, the RWs with the double-periodic background for two different elliptic modulus values $k=0.8$ and $k=0.9$ are shown in Figure \ref{fig3}(d). Here also, we observe some changes in the nature of the RWs. For example, the RWs amplitude decrease and also the orientation of the repeating wave patterns of the surrounding waves change as we enhance the $k$ value to $0.9$. From these observations, we conclude that the height and width of the RWs depends on the elliptic modulus ($k$) and system parameter ($\delta$) respectively. The outcome confirm that the parameters $k$ and $\delta$ control the RWs of Eq. (\ref{e1}). So these two parameters may be used to control the appearance of RWs from the experimental point of view. When replacing the eigenvalues $\lambda_2$ and $\lambda_3$ in the place of $\lambda_1$ we notice amplitude variations in each case. The other characteristics remained unchanged. 
	
	Three dimensional RW patterns ($|\hat{q}|^2$) on the double-periodic wave background (\ref{e5}) and their contour plots are demonstrated in Figure \ref{fig4} for  $k=0.9$ and $\delta=0.05$  using the generated RWs solution of Eq. (\ref{e1}). Figures \ref{fig4}(a)-\ref{fig4}(b) are plotted for $\lambda_1=\sqrt{\nu_1}$, whereas Figures \ref{fig4}(c)-\ref{fig4}(d) are plotted with $\lambda_2=\sqrt{\alpha+i\beta}$. RWs on the considered background and their contour plots for the eigenvalue $\lambda_3$ produce the same result (Figures \ref{fig4}(c)-\ref{fig4}(d)). This is because the third eigenvalue $\lambda_3$ is the complex conjugate of $\lambda_2$. So, we did not display the outcome obtained with $\lambda_3$ separately. In all the cases presented, we behold that the maximum amplitude of RWs is observed at the origin $(x,t)=(0,0)$. As in the previous case, the RWs have different amplitudes for each eigenvalue. The amplitude $|\hat{q}|^2$ is $17.846$ for $\lambda_1$ and it is $13.7307$ for $\lambda_2$. The amplitude of the RWs from the double-periodic wave Eq. (\ref{e5}) is larger compare to the one from the solution Eq. (\ref{e4}).
	
	Figure \ref{fig5} shows the two dimensional plots of RWs over a double-periodic wave solution ($\ref{e5}$) for $\lambda_1=\sqrt{\nu_1}$. When $x=0$, the cross sectional view with respect to $t$ is shown in Figure \ref{fig5}(a) whereas Figure \ref{fig5}(b) presents the cross sectional view with respect to $x$ when $t=0$, in both cases the value of $\delta$ is $0.05$. Figure \ref{fig5}(c) reveals the difference between RWs on the double-periodic wave solution of NLS (Eq. (\ref{e13}) with (\ref{e5}) when $\delta=0$) and the fifth-order NLS equations (Eq. (\ref{e13}) with (\ref{e5}) when $\delta=0.05$). Here, we notice that, the RW of a fifth-order NLS solution is very sharp and it narrows (orange color) when compared to the NLS solution (blue color). Similarly, the differences that occur in the solution Eq. (\ref{e13}) with (\ref{e5}) by varying the $k$ value is shown in Fig. \ref{fig5}(d). The RW amplitude decreases once we decrease the value of $k$ from $0.9$ (blue color) to $0.3$ (orange color). In both the Figures \ref{fig5}(c) and \ref{fig5}(d) we display the changes that occur in the orientation of the background waves by changing the values $\delta$ and $k$. The solution, Eq. (\ref{e13}) with (\ref{e5}) also gives the same observation when compared to the solution Eq. (\ref{e13}) with Eq. (\ref{e4}), that is the size and reduction in width that occur in the RW due to the system parameter. The height of the RW also decrease with respect to the elliptic modulus $k$. Here also the amplitude of the RWs and the surrounding periodic waves differ in each solution. We also observed that the background waves turn out to be larger when compare to Eq. (\ref{e13}) with Eq. (\ref{e4}). We also notice a similar outcome for the other eigenvalue $\lambda_2$. The only difference is the RW amplitude. 
	
	It has been shown that the creation of RWs are linked to the MI of the wave backgrounds \cite{na1,jms}. In the literature, several works have been reported on RWs and the MI of the background wave for the NLS family of equations \cite{chen4,bdec}. Motivated by the above studies, in the following section, we use the linear stability analysis to study the MI of the two different double-periodic waves and evaluate the rate of instability of two different double-periodic solutions of the fifth-order NLS Eq. (\ref{e1}).
	
	\section{Instability rate of the two different double-periodic waves of Eq. (\ref{e1})}
	In this section, we calculate the  growth rate for instability of both the double-periodic waves, Eqs. (\ref{e4}) and (\ref{e5}), of Eq. (\ref{e1}). To determine the perturbation relation for calculating the MI growth rate, we derive two (unperturbed  and perturbed) sets of linearized equations. First, we derive a set of linearized equations for the unperturbed part and then we derive the perturbed part equations. By comparing these two set of parts linearized equations we obtain the perturbation relation. To set up the unperturbed set of linearized equation we recast the Eqs. (\ref{e4}) and (\ref{e5}) in the following form
	\begin{align}
		q(x,t)=g(x,t)~e^{2i a t},\quad g(x+\omega_1,t)=g(x,t+\omega_2)=g(x,t),
		\label{e14}
	\end{align}
	in which the fundamental period in space and time coordinate are $\omega_1>0$ and $\omega_2>0$ respectively and one might consider $2a=1$ for Eq. (\ref{e4}) and $2a= k$ for Eq. (\ref{e5}).
	
	The solutions, Eqs. (\ref{e4}) and (\ref{e5}) have been rewritten in a generalized form given in Eq. (\ref{e14}), with $g(x,t)$ considered as the solution of the linear equation (\ref{s2}) with $g_1$ and $g_2$, which are given by
	\begin{align}
		g_1(x,t)=&\rho_1(x,t)~e^{\eta x+(\xi+i a ) t},\nonumber\\ g_2(x,t)=&\rho_2(x,t)~e^{\eta x+(\xi-i a ) t},
		\label{e20}
	\end{align}
	where $\eta$, $\xi$ are spectral parameters and $\rho=(\rho_1,\rho_2)^T$. Substituting Eq. (\ref{e20}) into (\ref{s2}), we obtain a system of equations, namely
	\begin{subequations}
		\label{e21}
		\begin{align}
			\rho_x+\eta \rho=\begin{pmatrix}
				\lambda & g\\
				-\bar{g} & -\lambda
			\end{pmatrix}\rho,
			\label{e22}\\
			\rho_t+\xi\rho=\begin{pmatrix}
				\hat{A}_1- ia & \hat{B}_1\\ 
				\hat{C}_1 & -\hat{A}_1+ ia
			\end{pmatrix}\rho,
			\label{e23}
		\end{align}
	\end{subequations}
	where,
	\begin{align}
		\hat{A}_1=\;&i\left(\lambda^2+\frac{|g|^2}{2}\right)+\delta(\bar{g}g_{xxx}-g\bar{g}_{xxx}+g_x\bar{g}_{xx}-\bar{g}_xg_{xx}+6|g|^2\bar{g}g_x-6|g|^2\bar{g}_xg)\nonumber\\&+2\lambda\delta(g\bar{g}_{xx}+\bar{g}g_{xx}-|g_x|^2+3|g|^4)-4\lambda^2\delta(g\bar{g}_x-\bar{g}g_x)+8\lambda^3\delta|g|^2+16\lambda^5\delta,\nonumber\\
		\hat{B}_1=&\;i\left(\lambda g+\frac{g_x}{2}\right)+\delta(g_{xxxx}+8|g|^2g_{xxx}+2g^2\bar{g}_{xx}+4|g_x|^2g+6\bar{g}g^2_x+6|g|^4g)\nonumber\\&+2\lambda\delta(g_{xxx}+6|g|^2g_x)+4\lambda^2\delta(g_{xx}+2|g|^2g)+8\lambda^3\delta g_x+16\lambda^4\delta g,\nonumber\\
		\hat{C}_1=&\;-i\left(\lambda\bar{g}-\frac{\bar{g}_x}{2}\right)-\delta(\bar{g}_{xxxx}+8|g|^2\bar{g}_{xx}+2\bar{g}^2g_{xx}+4|g_x|^2\bar{g}+6g\bar{g}^2_x+6|g|^4\bar{g})\nonumber\\&+2\lambda\delta(\bar{g}_{xxx}+6|g|^2\bar{g}_x)-4\lambda^2\delta(\bar{g}_{xx}+2|g|^2\bar{g})+8\lambda^3\delta\bar{g}_x-16\lambda^4\delta\bar{g},\nonumber
	\end{align}
	
	Upon perturbing the solution  (\ref{e14}) linearly,
	\begin{align}
		q(x,t)=g(x,t)e^{2iat}+p(x,t)~e^{2iat},
		\label{e15}
	\end{align}
	where $p(x,t)$ is a function of its arguments and substituting the latter expression into (\ref{e1}) and leaving out the quadratic and cubic terms in $p$, we arrive at
	\begin{subequations}
		\label{s224}
		\begin{align}
			\label{e16}
			&ip_t-2ap+\frac{1}{2}p_{xx}+2|g|p+g^2\bar{p}-i\delta(p_{xxxxx}+10(\bar{g}p+\bar{p}g)g_{xxx}+10|g|^2p_{xxx}\nonumber\\&+10(pg_x+gp_x )\bar{g}_{xx}+10(p\bar{g}_x+g\bar{p}_x)g_{xx}+20(\bar{p}g_x +\bar{g}p_x)g_{xx}+10gg_x\bar{p}_{xx}\nonumber\\&+10(g\bar{g}_x+2\bar{g}g_x)p_{xx}+20|g_x|^2p_x+10g_x^2\bar{p}_x+60(\bar{g}p+g\bar{p})|g|^2g_x+|g|^4p_x)=0.\\
			&i\bar{p}_t-2a\bar{p}+\frac{1}{2}\bar{p}_{xx}+2|g|\bar{p}+\bar{g}^2p+i\delta(\bar{p}_{xxxxx}+10(g\bar{p}+p\bar{g})\bar{g}_{xxx}+10|g|^2\bar{p}_{xxx}\nonumber\\&+10(\bar{p}\bar{g}_x+\bar{g}\bar{p}_x )g_{xx}+10(\bar{p}g_x+\bar{g}p_x)\bar{g}_{xx}+20(p\bar{g}_x +g\bar{p}_x)\bar{g}_{xx}+10\bar{g}\bar{g}_xp_{xx}\nonumber\\&+10(\bar{g}g_x+2g\bar{g}_x)\bar{p}_{xx}+20|g_x|^2\bar{p}_x+10\bar{g}_x^2p_x+60(g\bar{p}+\bar{g}p)|g|^2\bar{g}_x+|g|^4\bar{p}_x)=0.
		\end{align}
	\end{subequations}
	
	We consider a simple nontrivial separable solution of Eq. (\ref{s224}) in the form
	\begin{align}
		p(x,t)=n_1(x)~e^{ \Lambda t}, \quad \bar{p}(x,t)=n_2(x)~e^{\Lambda t},
		\label{e17}
	\end{align}
	where $n_1$ and $n_2$ are the functions of $x$  ($n(x)=(n_1(x),n_2(x))^{T}$) and $\Lambda$ denotes the spectral parameter. By using Eqs. (\ref{e16}) and (\ref{e17}) we rewrite the spectral stability problem as follows,
	\begin{align}
		\label{e18}
		i\Lambda\sigma_3 ~n+\begin{pmatrix} m_{11} & m_{21}\\ \bar{m}_{21} & \bar{m}_{11} \end{pmatrix}n=0, \quad
		\sigma_3 = \begin{pmatrix}
			1 & 0 \\
			0 & -1
		\end{pmatrix},
	\end{align}
	with
	\begin{subequations}
		\label{e19}
		\begin{align}
			m_{11}=&-2a+2|f|^2+\frac{1}{2}\partial_{xx}-i\delta(|g|^4\partial_x+60|g|^2\bar{g}g_x+20|g_x|^2\partial_x+20\bar{g}g_x\partial_{xx}\nonumber\\&+10g\bar{g}_x\partial_{xx}+20\bar{g}g_{xx}\partial_x+10\bar{g}_xg_{xx}+10g\bar{g}_{xx}+10g_x\bar{g}_{xx}+10|g|^2\partial_{xxx}\nonumber\\&+10\bar{g}g_{xxx}+\partial_{xxxxx}), \\ 
			m_{21}=~&g^2-i\delta(10gg_{xxx}+20g_xg_{xx}+10gg_{xx}\partial_x+10gg_{xx}\partial_{xx}+10g_x^2\partial_x\nonumber\\&+60|g|^2gg_x).
		\end{align}
	\end{subequations}
	Equation (\ref{e18}) is obtained through the perturbed solution. Now comparing Eq. (\ref{e18}) with Eq. (\ref{e23}) we could fix the bounded squared eigenvalues $\rho_1^2$ and $\rho_2^2$, as $n_1=\rho_1^2$, $n_2=-\rho_2^2$ and $\delta=i\delta$. We also obtain a relationship between the spectral parameters $\Lambda$ and $\xi$ in the form $\Lambda=2\xi$ by comparing the Eqs. (\ref{e18}) and (\ref{e23}). In another words, when substituting the obtained relations, $\delta$ and $\Lambda$, in Eq. (\ref{e18}) we arrive at Eq. (\ref{e23}). 
	
	The explicit expression for $\xi$ can be obtained from Eq. (\ref{e23}). The spectral problem or the coefficient matrix given in Eq. (\ref{e23}), provides a set of linear algebraic equations. A nonzero solution of these equations could be determined only when the determinant is zero for the coefficient matrix. This condition establish a connection between $\xi$ and $\lambda$, that is
	\begin{align}
		\xi^2+\tilde{W}(\lambda)=0,
		\label{e25}
	\end{align}
	with
	\begin{align}
		\tilde{W}(\lambda)=\lambda^4-2a\lambda^2+a^2+2b, \quad	|g_x|^2+|g|^4-4a|g|^2=8b.
		\label{e26}
	\end{align}
	\begin{figure*}[!ht]
		\begin{center}
			\begin{subfigure}{0.4\textwidth}
				\includegraphics[width=\linewidth]{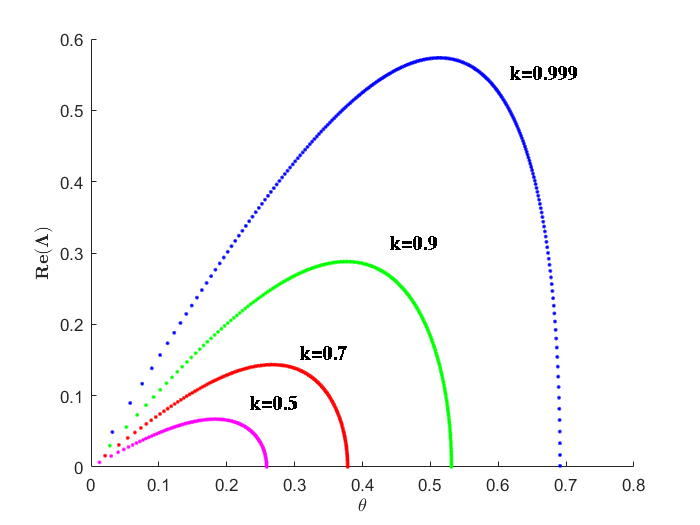}
				\caption{}
			\end{subfigure}
			\begin{subfigure}{0.4\textwidth}
				\includegraphics[width=\linewidth]{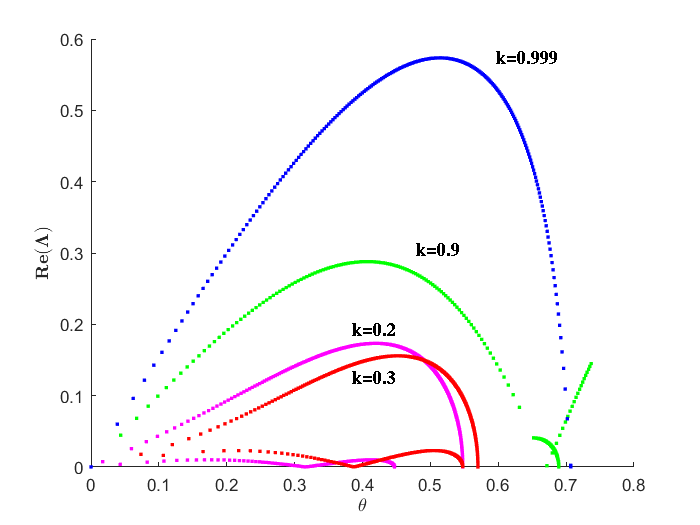}
				\caption{}
			\end{subfigure}
		\end{center}
		\vspace{-0.15cm}
		\caption{Instability rate for the double-periodic waves for different elliptic modulus values with $\delta=0.01$: (a) for Eq. (\ref{e4}), (blue, green, red, pink colors denotes the different values of $k$, which are $0.999$, $0.9$, $0.7$ and $0.5$ respectively) and (b) Instability rate for Eq. (\ref{e5}) (blue, green, red, pink colors denotes the different values of $k$, which are $0.999$, $0.9$, $0.3$ and $0.2$ respectively). }
		\label{fig6}
	\end{figure*}
	
	The polynomial $\tilde{W}(\lambda)$ could be rewritten as
	\begin{align}
		\tilde{W}(\lambda)=\lambda^4-\frac{1}{2}(g_1^2+g_2^2)\lambda^2+\frac{1}{16}(g_1^2-g_2^2),
		\label{e28}
	\end{align}
	with $4a=g_1^2+g_2^2$ and $8b=-g_1^2g_2^2$. The roots of the polynomial $\tilde{W}(\lambda)$ are $\pm\lambda_1$ and $\pm\lambda_2$. The polynomial $\tilde{W}(\lambda)$ have two positive roots which are
	\begin{align}
		\lambda_1=\frac{1}{2}(g_1+g_2), \quad \lambda_2=\frac{1}{2}(g_1-g_2).
		\label{e29}
	\end{align}
	Using Eq. (\ref{e29}) we rewrite Eq. (\ref{e28}) in the form
	\begin{align}
		\tilde{W}(\lambda)=(\lambda^2-\lambda_1^2)(\lambda^2-\lambda_2^2).
		\label{e30}
	\end{align}
	From Eq. (\ref{e28}), we get a relationship between $\xi$ and $\lambda_i$'s in the following form
	\begin{align}
		\xi=\pm i\sqrt{(\lambda^2-\lambda_1^2)(\lambda^2-\lambda_2^2)}
		\label{e32}
	\end{align}
	
	With the help of Eq. (\ref{e32}), we rewrite the value of $\Lambda$ in the form
	\begin{align}
		\Lambda=\pm 2i\sqrt{(\lambda^2-\lambda_1^2)(\lambda^2-\lambda_2^2)}\quad \text{or}\quad \Lambda=2\xi,
	\end{align}
	where the values of $\lambda$, could be obtained from the spectral problem (\ref{e22}), that is
	\begin{equation}
		\begin{pmatrix} \frac{d}{dx}+\eta & -g\\ -\bar{g} & -\frac{d}{dx}-\eta \end{pmatrix}\begin{pmatrix} \rho_1\\ \rho_2 \end{pmatrix}=\lambda \begin{pmatrix} \rho_1\\ \rho_2 \end{pmatrix},
		\label{e31}
	\end{equation}
	where $\eta=i\theta$ and the Floquet parameter, $\theta\in(0,\frac{\pi}{\omega_1})$. We solve Eq. (\ref{e31}), through numerically, using the eigenvalue solver technique \cite{chen1,dep}. Using the relation $\Lambda$, we can compute the instability rate of the considered system. In the numerical plotting part we consider the values of $\Lambda$ and $\delta$ are real.

	Instability rate of the two double-periodic waves (\ref{e4}) and (\ref{e5}) for the fifth-order NLS equation is illustrated in Figure \ref{fig6} with four different $k$ values with $\delta=0.01$. Here, $Re(\Lambda)$ is plotted against $\theta$, where the $\theta$ value is considered between the range $0$ to $\frac{\pi}{\omega_1}$. The highest instability rate of Eq. (\ref{e4}) is achieved when $k=0.999$ and this maximal instability rate decrease when the value of $k$ is decreased from $0.999$ to $0.5$ (Figure \ref{fig6}(a)). For $k=0$ the double-periodic solutions, Eqs. (\ref{e4}) and (\ref{e5}) converge to the NLS soliton and cnoidal wave, respectively and for $k=1$, the spectrum of both the solutions achieve Akhmediev breathers level. In Figure \ref{fig6}(b) we display the rate of instability for the double-periodic wave of Eq. (\ref{e5}). Here also the highest instability rate is obtained for $k=0.999$ ($k\approx1$). The instability rate decrease when $k$ value is decreased from higher ($0.999$) to lower ($0.3$). Further, when we decrease the value of elliptic modulus say $k=0.2$, the maximal instability suddenly increase but it is lesser than the $k=0.9$ curve. From Figure \ref{fig6}, we observe that both the double-periodic waves, the instability rate attained its maximal level to $0.999$ ($\approx 1$) and width of the spectrum also had the same range $0.7$. This is because, both the solutions converge to Akhmediev breathers when $k=1$. In both the solutions, the largest instability rate and the width of them gradually decrease when the system parameter ($\delta$) value is increased.
	\section{Conclusion}
	We investigated the RW solution with two different double-periodic wave backgrounds for the fifth-order NLS equation through DT. The approach of nonlinearization of the Lax pair or spectral problem was used to estimate eigenvalues and squared eigenfunctions associated with these double-periodic wave solutions. We constructed two different backgrounds that are periodic in both space and time by substituting the derived eigenvalues, squared eigenfunctions and the double-periodic seed solution into the DT (\ref{e3}). We then created the RWs over the constructed wave backgrounds by using the second linearly independent solution (which is of a non-periodic form and grows linearly on $x$ and $t$) to the Lax pair equation. We investigated the characteristics of RWs for various system parameter values $(\delta)$ and for the different eigenvalues. We observed that the orientation and the number of occurrences of repeating wave patterns of both the double-periodic wave changes in both space and time axis. We also noticed that the double-periodic waves and the RWs amplitude varies depending on its eigenvalues. To get a better understanding of these RWs over the double-periodic wave background, we drew two dimensional plots for two different values of system and elliptic modulus parameters. These figures brought out the differences between the RW structures in the basic NLS equation and its higher-order variant, say fifth-order NLS equation. In the case of higher-order NLS, the RWs are very sharp and it narrows inside the NLS RW due to higher-order dispersion effects. We also noticed that, by decreasing the value of $k$, the height of the RWs for both double-periodic wave solutions decreases. Hence, we concluded that the parameters $\delta$ and $k$ can be used to analyze the appearance/disappearance of RWs in several oceanographic and hydrodynamics experiments. We also calculated the growth rate for instability of both the double-periodic waves for the fifth-order NLS equation using the spectral stability analysis method. By perturbing the double-periodic wave solutions linearly and substituting this into the fifth-order NLS equation, we obtained a set of linearized equations. Upon solving this set of linearized equations, we obtained the instability relations. Finally, we computed the instability rate for a particular value of the system parameter and examined the rate of instability for four different values of  elliptic modulus parameter. Furthermore, RWs on a double-periodic wave background have a wide range of applications in mathematics and physics \cite{rand,kim}. The results of our paper may prove to be useful for diagnosing RWs on the surface of the ocean and understanding how random waves form, owing to MI \cite{ac1}. 
	\section*{Acknowledgements}
	NS wishes to thank MoE-RUSA $2.0$ Physical Sciences, Government of India for sponsoring a fellowship for this research work. MS acknowledges MoE-RUSA $2.0$ Physical Sciences, Government of India for sponsoring this research work. MS also acknowledges the research project sponsored by NBHM, Government of India, under the Grant No. $02011/20/2018$ NBHM(R.P)/R$\&$D II/$15064$.
	
\end{document}